%% file: main.tex
\documentclass[manuscript, nonacm]{acmart}
\AtBeginDocument{%
  \providecommand\BibTeX{{%
    \normalfont B\kern-0.5em{\scshape i\kern-0.25em b}\kern-0.8em\TeX}}}

\setcopyright{acmlicensed}
\copyrightyear{2018}
\acmYear{2018}
\acmDOI{XXXXXXX.XXXXXXX}

\acmConference[Conference acronym 'XX]{Make sure to enter the correct
  conference title from your rights confirmation emai}{June 03--05,
  2018}{Woodstock, NY}
\acmBooktitle{Woodstock '18: ACM Symposium on Neural Gaze Detection,
 June 03--05, 2018, Woodstock, NY} 
\acmISBN{978-1-4503-XXXX-X/18/06}

\definecolor{lg}{gray}{0.9}
\definecolor{red}{rgb}{.8, 0, 0}

\usepackage{tcolorbox}

\usepackage{tikz}

\newtcolorbox{example}[1]{colbacktitle=white,coltitle=black!75!white,size=small,fontupper=\small,title={#1}}

\newcommand*\circled[1]{\tikz[baseline=(char.base)]{
            \node[shape=circle,draw,inner sep=2pt] (char) {#1};}}

\begin{document}

\title{On the Opportunities of Large Language Models for Programming Process Data}

\author{John Edwards}
\email{john.edwards@usu.edu}
\affiliation{
  \institution{Utah State University}
  \streetaddress{4205 Old Main Hill}
  \city{Logan}
  \state{Utah}
  \postcode{84322-4205}
  \country{United States of America}
}

\author{Arto Hellas}
\email{arto.hellas@aalto.fi}
\affiliation{
  \institution{Aalto University}
  \city{Espoo}
  \state{Finland}
  \country{Finland}
}

\author{Juho Leinonen}
\email{juho.leinonen@iki.fi}
\affiliation{
  \institution{Aalto University}
  \city{Espoo}
  \state{Finland}
  \country{Finland}
}

\begin{abstract}
  \input{00-abstract}
\end{abstract}



\received{20 February 2007}
\received[revised]{12 March 2009}
\received[accepted]{5 June 2009}

\maketitle

\section{Introduction}

\input{10-introduction}

\section{Programming Process Data}
\label{sec:background}

\input{20-background}

\section{Possibilities of Large Language Models for Programming Process Data}
\label{sec:possibilities}

\input{30-possibilities-of-llms}

\section{Case Study: Analyzing Programming Process Data with LLMs}
\label{sec:case-study}

\input{40-case-study}

\section{Discussion}
\label{sec:discussion}

\input{50-discussion}

\bibliographystyle{ACM-Reference-Format}
\bibliography{main}






\end{document}

%% file: 00-abstract.tex
Computing educators and researchers have used programming process data to understand how programs are constructed and what sorts of problems students struggle with. Although such data shows promise for using it for feedback, fully automated programming process feedback systems have still been an under-explored area. The recent emergence of large language models (LLMs) have yielded additional opportunities for researchers in a wide variety of fields. LLMs are efficient at transforming content from one format to another, leveraging the body of knowledge they have been trained with in the process. In this article, we discuss opportunities of using LLMs for analyzing programming process data. To complement our discussion, we outline a case study where we have leveraged LLMs for automatically summarizing the programming process and for creating formative feedback on the programming process. Overall, our discussion and findings highlight that the computing education research and practice community is again one step closer to automating formative programming process-focused feedback.

%% file: 10-introduction.tex
Feedback can have a tremendous impact on learning and achievement~\cite{hattie2007power}. The level of detail of the feedback influences its effectiveness~\cite{wisniewski2020power}, and feedback can be given at many levels ranging from targeting how to work on and complete specific tasks to considering personal characteristics and behavior~\cite{hattie2007power,ott2016translating,keuning2018systematic}. In teaching and learning programming, automated assessment systems have been a key tool for providing feedback at a scale already for more than a half a century~\cite{hollingsworth1960automatic,paiva2022automated,keuning2018systematic}. Researchers have sought to automate step-by-step guidance~\cite{vihavainen2013scaffolding}, provide hints during the programming process~\cite{mcbroom2021survey}, improve programming error messages~\cite{becker2019compiler}, and aid in providing textual feedback by grouping similar code submissions together~\cite{nguyen2014codewebs,glassman2015overcode,koivisto2022evaluating}. 

To support the understanding of how novices construct programs, researchers and educators have been collecting increasing amounts of data from students' programming process~\cite{ihantola2015educational}. Such data can be collected at multiple granularities, ranging from final course assignment submissions to individual keystrokes from solving the assignments~\cite{ihantola2015educational}. Programming process data has been, for example, used to play back how students construct their programs step by step or keystroke by keystroke to create a broader understanding of the process~\cite{shrestha2022codeprocess, zhong2024insprog, heinonen2014using}. So far, despite shared efforts towards providing timely feedback to students~\cite{jeuring2022towards}, the potential of fine-grained programming process data for feedback purposes is still largely untapped. 

Large Language Models (LLMs) are a potential tool for realizing the transformation of programming process data into actionable feedback items. Within Computing Education Research, LLMs have broadened the horizon of what computing education researchers and practitioners can achieve~\cite{prather2023robots}, calling even for rethinking how computer science and programming is taught~\cite{denny2024computing}. Large Language Models have been shown to help in creating assignments~\cite{logacheva2024evaluating,sarsa2022automatic}, improve error messages~\cite{leinonen2023using,santos2023always}, fix students' code~\cite{koutcheme2023automated}, explain code~\cite{macneil2022generating,macneil2023experiences}, and respond to help requests~\cite{hellas2023exploring,koutcheme2024open}. At their core, LLMs are tools that allow transforming text-based content from one form to another, drawing on the data that they have been trained with in the process and the instructions provided by the user~\cite{vaswani2017attention,ouyang2022training}.

In this article, we discuss the potentials of LLMs for programming process data. We outline a case study of using LLMs for analyzing programming process data. In the case study, we first use LLMs for summarizing the programming process, followed by asking LLMs to provide feedback on the process. This article is structured as follows. In Section~\ref{sec:background}, we outline prior research on programming process data. Building on Section~\ref{sec:background}, Section~\ref{sec:possibilities} outlines our vision for the possibilities of LLMs for Programming Process Data. In Section~\ref{sec:case-study}, we outline our case study where we leveraged LLMs for analyzing programming process data. Finally, in Section~\ref{sec:discussion}, we summarize our discussion and outline possible future directions for research on using LLMs for programming process analytics.

%% file: 20-background.tex
In this section, we discuss the types and different uses of programming process data. Two surveys~\cite{ihantola2015educational,edwards2023review} are resources for discussion of both the different granularities of raw process data and the wide variety of studies and research questions being purused using this type of data.

\subsection{Types of data}
Programming process data has been collected in various forms. \citet{ihantola2015educational} discuss process data at six discrete levels and~\citet{edwards2023review} define four levels, which are, starting from the least granular: submissions and commits~\cite[e.g.,][]{ahadi2016number,koutcheme2022methodological}; executions, compilations, and file saves~\cite[e.g.,][]{carter2015normalized,jadud2006methods,kazerouni2017deveventtracker}; line-level edits, which capture all contiguous changes in a single line in a single event~\cite[e.g.,][]{brown2014blackbox,brown2018blackbox}; and finally, keystroke and character edits~\cite[e.g.,][]{leinonen2016automatic,leinonen2021fine,carter2017using,edwards2023review}. Our interest is in using occasional snapshots of code and analyze the progression of one snapshot to another. Any granularity of process data would potentially be suitable for our purposes, but, in practice, students may not submit or commit their data often enough, nor do executions or compilations guarantee sufficient temporal resolution. However, line-level and keystroke-level data are generally too fine-grained. In this paper, we use keystroke-level data but discretize it into code snapshots that immediately precede a break in keystrokes of at least five minutes.

The five-minute threshold is based on a probabilistic model by~\citet{hart2023accurate} that shows that breaks of five minutes statistically indicate a 50\% chance that the student has disengaged. Among the many other thresholds that have been used, which arbitrarily range from 60 seconds to one hour~\cite{price2016evaluation,leinonen2017comparison,murphy2009retina,kazerouni2017quantifying}, one study used a threshold of five minutes~\cite{leinonen2019exploring}, but it pre-dated Hart's statistical model and was arbitrarily chosen.

A small number of public datasets are available that would be suitable for our experiments, including Blackbox~\cite{brown2014blackbox,brown2018blackbox,brown2020blackbox}, which has line-level edit data, and a CSEDM challenge dataset\footnote{https://pslcdatashop.web.cmu.edu/Files?datasetId=3458} which has compile-level granularity. We chose the dataset provided by~\citet{edwards2023review}, which is keystroke-level, because it gives us more flexibility over the temporal resolution of code snapshots.

\subsection{Uses of programming process data}
Programming process data has been used in biometrics and authentication~\cite{morales2015keystroke,krishnamoorthy2018identification,leinonen2016automatic,longi2015identification,leinonen2016typing} and plagiarism detection~\cite{rodriguez2022tracking,hellas2017plagiarism,shrestha2022codeprocess}, but the most common use of programming process data is in predicting course outcomes. Early prediction work used features of compiler errors, including frequency and repetition~\cite{jadud2006methods,watson2013predicting,becker2016new,ahadi2015exploring,castro2017evaluating}. Later work incorporated other features derived from process data, including number of keystrokes~\cite{spacco2015analyzing}, number of IDE hints~\cite{estey2016can,estey2017automatically}, time-on-task~\cite{leinonen2022time}, and many others~\cite{edwards2023review}.

Pertinent to our work presented in this paper are studies that looked at how the code itself is constructed, though most of these use rudimentary measures. They include measuring incremental test writing and execution to derive how often a student works on their program~\cite{kazerouni2017deveventtracker}; clustering of students into tinkerers and planners~\cite{blikstein2014programming}; abstractly visualizing student progress toward a solution~\cite{piech2012modeling,shrestha2022codeprocess}; and tools to interactively visualize code snapshots to facilitate discussion~\cite{yan2019pensieve}. Intelligent Tutoring Systems (ITSs) have used various features for machine learning techniques to determine hints and feedback as students write code~\cite{barnes2008toward,rivers2017data,price2017evaluation,price2017isnap}. In general, these approaches use canonical solutions to train the models and often find a student's code snapshot in a model code progression to determine what a suggested next step might be. While superficially similar to our approach, usage of our system differs in two fundamental ways. The first is that ITSs generally suggest a next step, whereas our feedback is potentially retrospective. Second, we are interested in both step-wise and holistic feedback. The idea of commentary on the overall arc of developing a program is not a prominent concept in ITSs.

Programming process data is a largely untapped resource in understanding how students write code. This is recognized by researchers who have proposed questions that could be answered given a suitable feature set and analysis techniques.~\citet{shrestha2022codeprocess} asked what exactly good programming process is. Is it student dependent? Is it dependent on the type of problem? If so, how?~\citet{edwards2023review} asked what, exactly, flailing and learning looks like. Some work on this has been done~\cite{spacco2015analyzing} using time spent on exercises and submission frequency as features. We suggest that richer features would lead to deeper understanding of student cognition and behavior.

%% file: 30-possibilities-of-llms.tex
Large language models provide many exciting opportunities for analyzing and utilizing programming process data. One potential application of LLMs is to summarize the process into natural language. Before LLMs, most work in programming process data used derived metrics like time-on-task, number of events, etc. to analyze the process~\cite{ihantola2015educational,leinonen2019keystroke}, but here the utility of the data is only as good as the metrics derived from it. With LLMs, more abstract aspects of the process could potentially be derived. Another stream of work has looked into visualizing the programming process based on programming process data~\cite{xie2023developing,ditton2021external,shrestha2022codeprocess,heinonen2014using}, but here one issue is that the visualizations might be hard to interpret, or if the whole process is visualized keystroke-by-keystroke (see e.g.~\cite{xie2023developing,ditton2021external,heinonen2014using}), it is very time consuming to analyze.

Another potential application of LLMs for programming process data is to give feedback to students. When the whole process is available, feedback could be given with different granularities. For example, it could be possible to give high-level feedback on the whole process (e.g., ``\textit{you should run code more frequently}'') or low-level feedback that considers the process (e.g., ``\textit{you recently modified your function `calculateSum' but there is a small bug}''). Ideally, feedback using LLMs could be given \textit{during} the process, which could support students while they are working on their programs.

Some recent work has looked into generating synthetic data using LLMs~\cite{hamalainen2023evaluating, park2022social, park2023generative}, which is one potential application of LLMs for programming process data. One issue with programming process data, especially very fine-grained data such as keystroke data, is that individuals can sometimes be identified using ``keystroke dynamics'', i.e., based on their unique typing rhythms~\cite{longi2015identification,teh2013survey,ali2017keystroke}. This is an issue for sharing data openly~\cite{leinonen2017preventing,edwards2023review}, since typically data that contains student personal information or that can be traced back to the student cannot be shared. In Europe, the General Data Protection Regulation (GDPR) specifically lists biometric data (which keystroke data falls under~\cite{teh2013survey}) under ``special categories of personal data'' which have more strict requirements under the GDPR. While some work has looked into the de-identification of keystroke data~\cite{leinonen2017preventing,edwards2023review}, de-identification can reduce the utility of the data~\cite{leinonen2017preventing}. Thus, generating synthetic programming process data from authentic programming process data utilizing LLMs could help advance research, as long as the generated data is equal to real data quality-wise.

One challenge in utilizing process data with LLMs is the cost of using LLMs, which for many commercial LLMs is based on the number of tokens used\footnote{See, for example, OpenAI's pricing: \url{https://openai.com/api/pricing/} (Retrieved 4 June 2024).}. As process data typically is large~\cite{ihantola2015educational,edwards2023review}, the cost could be great. However, as the context windows of LLMs continue to increase\footnote{For example, Google's Gemini has a context window of up to ten million tokens: \url{https://blog.google/technology/ai/google-gemini-next-generation-model-february-2024/\#context-window} (Retrieved 4 June 2024).}, and the capabilities of open-source models increase, the costs of using LLMs might reduce.

%% file: 40-case-study.tex
Here, we outline our case study of analyzing programming process data using LLMs. For the case study, we explored how well three different LLMs 1) summarize the programming process based on the programming process data and 2) generate programming process feedback for students based on the programming process data.

\subsection{Process data and preprocessing}

For the process data, we used a publicly available Python dataset published by Edwards et al.~\cite{edwards2023review}. The dataset has 44 participants and 8 assignments from an introductory programming course organized at Utah State University in the United States. The dataset was collected using the PyPhanon plugin~\cite{edwards2022practical} for PyCharm\footnote{\url{https://www.jetbrains.com/pycharm/}}, and it consists of 1 million unique events. For the present work, we focused on three of the assignments. Outlines of the assignments are given in Table~\ref{tbl:assignments} and are taken from assignment descriptions included with the dataset. 

\begin{table}[]
    \caption{Summary descriptions of the three assignments included in the case study.\label{tbl:assignments}}    \centering
    \begin{tabular}{l|p{12cm}}
    \toprule
         Assignment name & Brief description \\
         \midrule
        Fluky numbers & Calculate fluky numbers. ``You will write a program that will find Fluky Numbers. A Fluky Number is a number that is equal to a random number, where the random number is the first random number generated after seeding a random number generator with the sum of all factors of a number, not including itself.'' \\
        Zookeeper & Zookeeper and elephants in pens. ``Your job is to write a program to simulate [a zookeeper checking elephant pens] 100,000 times to determine [the probabilities of the zookeeper finding an elephant in each pen].'' \\
        Number pyramids & Print a number pyramid. ``The user will enter the number of rows to have in a pyramid of numbers. Based on the user's input the program will display a pyramid where each row contains the number $x$, which is printed $x$ times, where $x$ is the row number.'' \\
        \bottomrule
    \end{tabular}
\end{table}

Using the logs from five students working on three assignments, we construed the programming process from the logs into a sequence of state snapshots that outlined how the assignment code evolved over time from the beginning to the end. For the present analysis, we created the sequence of state snapshots by including the first and last code states and each code snapshot immediately preceding breaks of five minutes or more. Table~\ref{tbl:state_descriptions} provides descriptive statistics on the number of state snapshots for each of the students for the included assignments.

\begin{table}[]
    \caption{Summary of number of states in the programming process data included in the data for LLMs for analysis.\label{tbl:state_descriptions}}    \centering
    \begin{tabular}{l|r|r|r|r}
    \toprule
         Student & Average states & Fluky numbers states & Zookeeper states & Number pyramid states \\
         \midrule
         S1 & 9.3 & 16 & 8 & 5 \\
         S2 & 8 & 11 & 6 & 7 \\
         S3 & 7.7 & 10 & 6 & 7 \\
         S4 & 7.7 & 9 & 3 & 11 \\
         S5 & 5.3 & 6 & 4 & 6 \\
         \midrule
         Avg. & 7.6 & 10.4 & 5.4 & 7.2 \\
    \end{tabular}
\end{table}

\subsection{Large language models and prompt engineering}

For the analysis, we used Anthropic's Claude 3 Opus\footnote{\url{https://www.anthropic.com/news/claude-3-family}, version \texttt{claude-3-opus-20240229}}, OpenAI's GPT-4 Turbo\footnote{\url{https://platform.openai.com/docs/models/gpt-4-and-gpt-4-turbo}, version \texttt{gpt-4-0125-preview}}, and Meta's LLaMa2 70B Chat model\footnote{\url{https://llama.meta.com/llama2}, HuggingFace version \texttt{meta-llama/Llama-2-70b-chat-hf}}. The first two models are proprietary, while the third model is an open-source model. The models were selected to represent a snapshot of the state-of-the-art LLMs at the time of the analysis in March and April of 2024. We acknowledge that OpenAI has the GPT-4 model that is known to perform better than the GPT-4 Turbo. However, the context window size for the GPT-4 version that was available for us was limited to 8k tokens, and our analyses highlighted that this was not sufficient. Thus, we opted for the GPT-4 Turbo which has a context window size of 128k tokens. This is also a limitation in some of our analyses where we leverage the LLaMa2 70B Chat model. 

Our prompt engineering had the objective of having LLMs describe the process a student took to write their computer program and then to provide feedback to the student as if the AI were an expert teaching assistant (TA). Following prompting best practices~\cite{ekin2023prompt}, we iteratively explored and refined a range of prompts, including providing context and domain information, personalizing the responses, providing information about the format of the input data, and providing guidelines and constraints on the expected outcomes. One of the authors was in charge of the prompt engineering process, sharing intermediate results and observations with the other authors. During the prompt engineering process, the research team met weekly, commenting on the outputs and the prompts, and discussing further prompt improvement possibilities and briefly evaluating their impact on the outcomes. 

During the prompt engineering process, we observed that the models were relatively poor at incorporating timestamp information. As an example, if a student had a break that consisted of multiple days, an LLM might suggest that the student was thinking hard about the problem for days. Thus, we omitted the timestamp data. We further explored specifying and not specifying the number of items in the feedback, and in the end resorted to not restricting the number of items for the present evaluation.

\input{prompts/2-process-feedback-prompt}

The final prompt for providing feedback on the programming process is outlined in Figure~\ref{fig:process_feedback_improvement_prompt}. When comparing the prompt for providing feedback on the programming process to the prompt for summarizing the programming process, the main differences are in the persona ``[...] and an expert in summarizing how students construct their programs [...]'' and the explicit task instructions ``As an extremely good introductory programming teaching assistant, summarize how the student constructed their program based on the following programming process data. Respond as if you were talking to the student. Only summarize the programming process data. Do not provide any suggestions or feedback.''.

\subsection{Data generation}

To generate the data, we used Anthropic's and OpenAI's APIs for Claude 3 Opus and GPT-4 Turbo respectively, while LLaMa2 70B Chat model was run on the HuggingFace platform\footnote{\url{https://huggingface.co/}}. The data generation led to a total of 45 programming process descriptions (3 assignments $\times$ 5 students $\times$ 3 models) and 45 feedbacks based on the programming process, leading to 90 entries. Table~\ref{tbl:data_generation_summary} provides descriptive statistics of the data generation process, outlining the average response time and the average response length for each of the models.

\begin{table}[]
    \caption{The average response time and the average response length for each of the models.\label{tbl:data_generation_summary}}    \centering
    \begin{tabular}{l|r|r}
    \toprule
         Model & Average response time (seconds) & Average response length (characters) \\
         \midrule
         Claude 3 Opus   & 28.6 & 1814 \\
         GPT-4 Turbo     & 30.2 & 3086 \\
         LLaMa2 70B Chat & 14.2 & 1943 \\
    \end{tabular}
\end{table}

\subsection{Evaluation}

The evaluation was divided into three parts. First, one of the researchers conducted a surface-level summary evaluation of the outputs to study whether the models followed the task. This was followed by three researchers analyzing a subset of the data to form a shared understanding of the data. Finally, one of the reviewers conducted a thematic analysis of the improvement suggestions in the feedback to identify recurring themes in the outputs.

\subsection{Results}

\subsubsection{Surface-level analysis}

On a surface level, the LLMs followed the instructions and provided expected outputs adequately. For the 15 inputs (5 students and 3 assignments), GPT-4 Turbo generated 15 acceptable (see below for our definition of acceptable) programming process summaries and 14 feedbacks, Claude 3 Opus generated 13 acceptable summaries and 11 feedbacks, and LLaMa2 70B Chat generated 10 acceptable summaries and 7 feedbacks. While GPT-4 Turbo and Claude 3 Opus provided a response for all student-exercise-prompt combinations, LLaMa 70B Chat failed to provide a response in two instances (one summary and one feedback) due to the prompt being too long. 

The main reason why a summary was not deemed acceptable was that the response focused on describing one of the code states but did not summarize the process. Similarly, the main reason when a programming process feedback was not deemed acceptable was that the feedback did not explicitly include information about the process, but provided more generic improvement suggestions such as suggesting more efficient ways to implement the algorithms, adding type hints, and so on.

\subsubsection{Analysis of programming process improvement suggestions}

Overall, while the models did often provide similar suggestions, there were also differences in the focus and style of the feedback. The GPT-4 Turbo model often used a logical structure to the feedback consisting of generic comments, specific comments, and a conclusion. The feedbacks from Claude 3 Opus and LLaMa 70B Chat model on the other hand were typically more list-like, providing suggestions directly. When considering the programming process feedback as a whole, the models typically followed the implicit instruction of being constructive and providing feedback in a positive fashion as the feedback was directed to students. The feedback items often included something that the student was praised on, even if just for adding comments or for exploring, as shown in the following quotes (mainly from GPT-4 Turbo).

\begin{quote}
\textit{``It's great to see you organizing your code with comments, making it easier to understand the purpose of each section. Keep maintaining this practice.''}
\end{quote}

\begin{quote}
\textit{``It's great that you started by experimenting with the random number generation in steps 1 and 3 to understand how it works. This exploratory coding is a good way to familiarize yourself with new concepts.''}
\end{quote}

The models also often captured back and forth movement where students first added code and then removed it, as well as scenarios where students started working on implementations that were not in line with the problem statement. Such feedbacks included comments that would suggest the student to step away from coding to thinking about the problem, as shown by the following quotes.

\begin{quote}
\textit{``The initial steps indicate a bit of confusion regarding the problem statement. [...] Taking time to fully understand the problem before starting to code can save a lot of time and confusion.''}    
\end{quote}

\begin{quote}
\textit{``You seem to jump into implementing parts of the solution without fully forming the algorithm in your mind or on paper first. It's important to sketch out a basic algorithm or pseudocode before you begin coding, especially for a problem with several specific steps like this one.''}    
\end{quote}

The feedbacks also at times referenced to the assignment handout that was included in the prompt, as shown in the next quote. The handout for the ``Fluky numbers'' assignment had explicit hints on how to efficiently solve the problem, and the student had followed the hint in their implementation.

\begin{quote}
\textit{``You had the right idea in step 008 to keep track of the number of Fluky Numbers found and to terminate the search after finding seven. This directly aligns with the hint provided and is crucial for optimizing performance.''}    
\end{quote}

The feedbacks also included more common aspects that could be also non-assignment -specific. These suggestions included incremental implementation and testing of the solution, planning ahead, reducing unnecessary outputs and comments, removing commented out code, adding comments to code to improve readability, dividing code into functions, and other aspects such as variable and function naming. The following example starts by orienting the student with specific references to code snapshots and then uses that to make the general suggestion of not leaving commented-out code in the final submission. It then explains why the student should follow the suggestion (for code readability).

\begin{quote}
\textit{``As you iterated from Step 002 to Step 005, it looks like you attempted to address the issue of centering the pyramid but struggled to find the right solution on your first tries. This is completely normal and part of the learning process. However, it's important to not leave commented-out code or unused variables in your final submission, as it can make the code harder to read. Find the solution that works best and then refactor your code to remove any unnecessary parts.''}
\end{quote}

In one submission, the student used an obscenity in a comment. The feedback included this:

\begin{quote}
\textit{``In Step 001, you've used an unfortunate typo, "fu**tion", which seems to be an attempt at humor but can come across as unprofessional, especially in a learning or professional setting. In the future, I recommend keeping comments and code professional and clear. This will not only help maintain a positive learning environment but also prepare you for professional coding practices.''}
\end{quote}

The LLMs in the context of our study were not immune to hallucinations. For example, the following feedback recommends one function instead of another for performance reasons when, in fact, \texttt{random.randint(0, k)} is an alias for \texttt{random.randrange(0, k+1)}:

\begin{quote}
\textit{``The code uses the `random.randint(0, k)` function to generate a random number between 0 and `k`. However, this function can be slow for large values of `k`. Instead, use the `random.randrange(0, k+1)` function, which is faster and more efficient.''}
\end{quote}

\subsubsection{Researcher insights from programming process summarization and feedback}

The AI systems were explicitly instructed not to include feedback on the code itself, but rather to focus on the process of writing the code. (One round of prompting includes the statement, "Only provide feedback on how to improve the programming process. Do not provide any summary description of the programming process data and do not provide explicit suggestions on how to improve the code." Other prompts include similar statements.) However, the LLMs were found to have great difficulty in recognizing the difference between the process a student takes to write code and the code itself. For example, the AI gave the following process-related feedback to one student:

\begin{quote}
\textit{``Incremental Development and Testing: I noticed that you began writing large blocks of code before testing smaller units. It's important to approach complex problems by breaking them down into smaller, more manageable pieces. Once you write a small part, such as a function to calculate the sum of factors, test it thoroughly before moving on. This will help you catch errors early and understand each part of your code better. Consider implementing your code piece by piece, testing each function independently before integrating it into your main logic.''}
\end{quote}

The same feedback set also includes the following, which relates to the code itself:

\begin{quote}
    \textit{``Efficiency in Your Loops and Functions: Your approach to finding the Fluky Numbers involves quite a bit of redundancy and potentially unnecessary calculations. For example, in calculating the sum of factors, think about how you might optimize this process to avoid redundant calculations or checks. In addition, I see you've modified the `factorSum` function to return both a sum and a count of factors, but the count doesn't appear to be utilized in later logic. Always review your functions to ensure they are doing exactly what is needed for the task at hand, no more, no less. This will help keep your code efficient and focused.''}
\end{quote}

The fact that the feedback includes the word ``process'' in the feedback indicates that it may be confusing the process of code execution with the process of code writing. Every single summary and feedback output from the LLMs included something relating to the code and not process. In retrospect, this should not have been surprising to us. Even we humans tend to find it challenging to distinguish between the two. One possible reason for this is that software engineers and programming educators are attuned to thinking about code but think less about the evolution of code into its final form. Thus, when we look at a code snapshot, we tend to analyze it as a standalone product, rather than in its temporal context. Furthermore, considering it through the lens of cognitive load theory, analyzing a code snapshot relative to a prior snapshot requires roughly double the working memory, which is already stretched when reading code. From a philosophical perspective, it makes sense that LLMs suffer from this as well, as they are trained on data generally generated by humans, and there is very little data that deals with the programming process.

Furthermore, the majority of generated summaries and feedback that are related to the programming process are generic. The above example that suggests that the student use incremental development and testing is typical: incremental development is so commonplace that the only thing keeping it from becoming cliche is that it is so rarely actually practiced. Whereas our vocabulary of things that can be improved in code is vast (e.g., variable names, comments, duplicated code, algorithm complexity, code organization into functions, etc), educators have far fewer candidate improvements to the programming process, which are generally limited to incremental development, early decomposition, and planning ahead.

A challenge in evaluating quality of summaries and feedback is their subjectivity. Feedback in particular, is a challenge because, in some cases, educators disagree on best practices. For example, many educators encourage up-front design whereas others encourage a more agile, refactoring approach. In order to make evaluation of AI output objective, the AI would need to be primed with opinions on controversial topics. Related to this, educators would need to decide which topics are the most important. In our case study, the AI responses were often excessively long, so the educator would need to limit the length of output and indicate to the AI which topics to prioritize.

So what criteria should an AI be judged on? We propose five scores. The first would be a hallucination score. This would be straightforward to measure with standard multi-rater coding practices. The second score would measure how well the AI distinguishes between process and analysis of static code. As discussed, this distinction is challenging for both LLMs and humans. However, we found that determining how well the AI makes the distinction is actually not too difficult. The human evaluator needn't consult the code snapshots to determine if the LLM output addresses process or code. The third score would be specificity against genericness relative to the given submission. An AI that speaks in tropes is not useful. And finally, scores on correctness and utility would be needed. These are straightforward measures to reason about but are actually the most difficult to evaluate. It requires the evaluator to consider code snapshots relative to other snapshots which, as discussed above, is challenging.

One last note: we found that the AI systems distinguish extremely well between generating summaries and generating feedback. That is, when we ask the LLM to give a summary, the output never focuses on feedback, and vice versa.

%% file: prompts/2-process-feedback-prompt.tex
\begin{figure}[h!tbp]
\centering
\begin{example}{Prompt format for providing feedback on the programming process}

You are an introductory programming teaching assistant who is an expert in analyzing programming process data and an expert in providing suggestions on improving the programming process based on the programming process data \circled{1}. You are studying how a student solved a programming problem by analyzing the programming process data \circled{2}. \\

Here's the handout for the programming problem:\\

\textcolor{blue}{<handout>} \circled{3} \\

The programming process data is described as a sequence of steps, where each step is numbered and each step has the assignment header and the code. The format of the data is as follows: \circled{4} \\

\#\#\#\#\# \\
Step: step identifier \\
\#\#\#\#\# \\
(the code state that the student had at this step) \\

As an extremely good introductory programming teaching assistant, provide suggestions on how to improve the programming process based on the following programming process data. Respond as if you were talking to the student. Only provide improvement suggestions based on the data. \circled{5} \\

\textcolor{blue}{<process data>} \circled{6}

\end{example}
\caption{Prompt format for providing feedback on the programming process. The prompt included (1) a description of the context and the persona, (2) the broader task, (3) the handout, (4) the format of the input data, (5) the explicit task instructions, and (6) the process data.}
\label{fig:process_feedback_improvement_prompt}
\end{figure}

%% file: 50-discussion.tex
Overall, we here make an argument that exploring the ability of LLMs to provide feedback on programming process is worthwhile. The argument is in two parts. The first is the argument that feedback itself on programming process is useful. While giving students feedback on code is common, giving them feedback on how they went about writing the code is rare. This is because process data isn't easy to collect, but also because little research has been done into best practices in code evolution in the context of novice programmers. We claim that guiding students on this front will result in less frustration, better learning, reduced attrition, and increased diversity in the computer programming community. If we accept that providing guidance to students on process is useful, then we can address the other part of the argument, that AI in general, and LLMs specifically, are well-suited to providing this feedback to students. The primary reason for this is that analyzing how a code evolved is far more difficult than analyzing a single code snapshot. Well-prompted LLMs have vast memory resources and can evaluate far more code submissions than a human, and can do it in a short amount of time.

One way to make process feedback even more effective could be to combine it with a playback tool such as \textit{KeystrokeExplorer}~\cite{edwards2023review}. We note that in our case study, we considered only snapshots of code immediately preceding a break of at least five minutes. During our analysis, we found that this granularity often left the reader disoriented. Actually watching the keystroke-level evolution of the code between snapshots could be helpful in keeping the student oriented as they receive the AI-generated feedback, and also potentially create self-reflection opportunities, in a similar way as one could e.g. review their chess games.

We see that already finding out the appropriate granularity (and data format) could significantly improve the preliminary results outlined in this article, not to mention the emergence of newer and more powerful LLMs. We also see that using LLMs for creating feedback on the programming process data and providing that feedback to students during the programming process could help in pinning down the points in time when feedback should be given, which has been an open question in prior programming process feedback research~\cite{jeuring2022towards}. To begin the discussion at the conference, and beyond, we outline the following five research directions for using LLMs on programming process data:

\begin{itemize}
    \item Cost-efficient representations of programming process data for LLMs, including evaluating data representation types such as diffs and exploring the utility of different data granularities.
    \item Providing insights from programming process data to students, exploring when and how to present the insights for maximal effectiveness, including incorporating the LLM-generated insights into existing systems such as IDEs and programming process visualizers.
    \item Analyzing the subjectivity of LLM-generated insights and their utility to learners and instructors, and identifying contextual factors that contribute to the subjectivity.
    \item Exploring the opportunities of leveraging programming process data with LLMs for supporting students in learning about the process of programming, which has been classically highlighted as one of the challenges in learning complex skills such as programming~\cite{bennedsen2005revealing,collins1991cognitive,vihavainen2011extreme}. 
    \item Training and fine-tuning LLMs with processing programming process data for more efficient and higher quality programming process insights.
\end{itemize}